\documentclass[aps,prl,twocolumn,floats,floatfix,english]{revtex4}
\usepackage[T1]{fontenc}
\usepackage[latin1]{inputenc}
\usepackage{amsmath}
\usepackage{amssymb}
\usepackage{setspace}
\usepackage{epsfig}

\makeatletter

\providecommand{\LyX}{L\kern-.1667em\lower.25em\hbox{Y}\kern-.125emX\@}

\usepackage{babel}
\makeatother
\begin{document}

\title{Clustering and information in correlation based financial networks}

\author{J.-P. Onnela$¹$}
\email[]{jonnela@lce.hut.fi}
\author{K. Kaski$¹$}
\author{J. Kertész$^{1,2}$}

\affiliation{$¹$Laboratory of Computational Engineering, Helsinki University of Technology, \\ 
P. O. Box 9203, FIN-02015 HUT, Finland}

\affiliation{$^2$Department of Theoretical Physics, Budapest University of Technology and Economics,\\
Budafoki út 8, H-1111, Budapest, Hungary.}

\begin{abstract}
Networks of companies can be constructed by using return correlations. A crucial issue in this approach is to select the relevant  correlations from the correlation matrix. In order to study this  problem, we start from an empty graph with no edges where the  vertices correspond to stocks. Then, one by one, we insert edges  between the vertices according to the rank of their correlation  strength, resulting in a network called asset graph. We study its  properties, such as topologically different growth types, number and  size of clusters and clustering coefficient. These properties, calculated from empirical data, are compared against  those of a random graph. The growth of the graph can be classified  according to the topological role of the newly inserted edge. We find  that the type of growth which is responsible for creating cycles in  the graph sets in much earlier for the empirical asset graph than for the random graph, and thus  reflects the high degree of networking present in the market. We  also find the number of clusters in the random graph to be one order  of magnitude higher than for the asset graph. At a critical  threshold, the random graph undergoes a radical change in topology  related to percolation transition and forms a single giant  cluster, a phenomenon which is not observed for the asset graph. Differences  in mean clustering coefficient lead us to conclude that most information  is contained roughly within 10\% of the edges.
\end{abstract}

\maketitle

\section{Introduction}

In a financial market the performance of a company is compactly
characterised by a single number, namely the stock price. This is
thought to be based on available information, although it is heavily
debated what information it should reflect.  In the world of business
and finance, companies interact with one another, creating an evolving
complex system \cite{santafe}.  Although the exact nature of these
interactions is not known, as far as price changes are concerned, it
seems safe to assume that they are reflected in the equal-time
correlations.  These are central in investment theory and risk
management, and also serve as inputs to the portfolio optimisation
problem in the classic Markowitz portfolio theory\cite{Mark}.

Network theory \cite{barabasi} provides an approach to complex
systems with many interacting units where the details of the
interactions are of lesser importance, it is their bare existence
what is focused on. Recently this approach has proved to be extremely
useful in a broad field of applications ranging from the Internet to
microbiology. Obviously, the economy is a good hunting field to search for
networks. \cite{caldarelli} 

In this paper we study a financial network where the vertices
correspond to stocks and the edges between them to distances, which
are transformed correlation coefficients. Mantegna was the first \cite{Man1} 
to construct networks based on stock price correlations
and the idea was followed by a series of papers \cite {kull_ker_man,
short, long, bali, Gio2, Gio}. Recently, also time-dependent
correlations were studied, resulting in a network of influence \cite{Kas}. 
Here we deal with a network, which we have termed asset graph
and introduced in \cite{intia}. It is a natural extension to our
previous work with asset trees \cite{short,long,bali}, based on the
idea by Mantegna \cite{Man1}.

We focus on the construction and clustering of the asset graph.  We
would like to emphasise that the important issue of information versus noise
is closely related to our study. Although the estimated correlation
matrix is a simple measure of coupling between stocks, it suffers from
similar problems as the stock price on which it is based; due to a
considerable degree of noise its information content is questionable.
The general problem with empirical data is that the correlation matrix
of $N$ assets is determined from $N$ time series of length $T$, and if
$T$ is not very large compared to $N$, one should expect the resulting
empirical correlation matrix to be dominated by measurement noise.
The fact that a certain part of the asset tree is robust,
i.e. changes very slowly in crash free times \cite {short, long} already 
points towards the existence of an information core. Here we would
like to explore this issue further.

The problem of information content of the correlation matrix is
central to portfolio theory. There have been several attempts to
analyse this issue.  One is based on the random matrix theory, which
offers an interesting comparative perspective \cite{Mehta}.
The idea is that the properties of an empirical correlation matrix are
compared to a null hypothesis of purely random matrix as can be
obtained from a finite time series of strictly independent assets. It
is postulated that deviations from the theoretical predictions are
indicative of true information.  The general finding is that empirical
correlation matrices are dominated by noise \cite{Lal,Ple}. There have
also been simulation-based approaches to study the effect of time
series finiteness \cite{Paf}, where the use of artificial data enables
isolation of errors due to sources other than finite $T$. A different
but intimately related approach has been preferred in the finance
literature, namely the principal component analysis
\cite{principal}. Recently the independent component analysis, a
different tool of multivariate statistical analysis has also been
applied to such problems \cite{indep}. 

We would like to follow a more geometrical alternative, based
on financial networks, which gives rise to an interesting parallelism with the
previous line of work.  Just as random matrix theory yields a
benchmark by establishing a null hypothesis of a totally random
matrix, random graph theory establishes a null hypothesis of a totally
random graph. In other words, one can compare the results obtained for
empirical graphs against those of random graphs, which are well known
\cite{bollobas}, and interpret deviations from random behaviour as
information.

The paper is organised as follows. In Section 2 we recapitulate the
methodology for constructing asset trees and asset graphs.  In Section
3 we study their differences 
due the clustering observed in the asset graph but not in asset tree.
In Section 4 we explore a sample asset graph further, and compare the
results to a random graph.  At the end of the section we briefly discuss
the problem of noise versus information in the light of our results. Finally, we summarise 
the results of the paper in Section 5.

\section{Methodology for constructing asset graphs and asset trees}

Earlier we have studied the time evolution of asset trees in
\cite{short, long, bali} and extended our approach to asset graphs in
\cite{intia}, where the two approaches were explicated and
compared. Let us first recapitulate the two methodologies. Consider
a price time series for a set of $N$ stocks and denote the closure price
of stock $i$ at time $\tau$ (an actual date) by $P_{i}(\tau)$, and
define the logarithmic return of stock $i$ as $r_{i}(\tau)=\ln
P_{i}(\tau)-\ln P_{i}(\tau-1)$. We extract a time window of width $T$,
measured in days and in this paper set to $T=1000$ (equal to four
years, assuming 250 trading days a year), and obtain a return vector
$\boldsymbol r_{i}^t$ for stock $i$, where the superscript $t$
enumerates the time window under consideration. Then equal time
correlation coefficients between assets $i$ and $j$ can be written as

\begin{equation}
\rho _{ij}^t=\frac{\langle \boldsymbol r_{i}^t \boldsymbol r_{j}^t
  \rangle -\langle \boldsymbol r_{i}^t \rangle \langle \boldsymbol
  r_{j}^t \rangle }{\sqrt{[\langle {\boldsymbol r_{i}^t}^{2} \rangle
      -\langle \boldsymbol r_{i}^t\rangle ^{2}][\langle {\boldsymbol
        r_{j}^t}^{2} \rangle -\langle \boldsymbol r_{j}^t \rangle
      ^{2}]}}, 
\end{equation}

\noindent where $\left\langle ...\right\rangle $ indicates a time
average over the consecutive trading days included in the return
vectors. These correlation coefficients between $N$ assets form a
symmetric $N\times N$ correlation matrix $\mathbf{C}^t$. The different time
windows are displaced by $\delta T$, where we have used a step size of
one month, i.e., $\delta T = 250/12 \approx 21$ days, which gives rise
to interpreting the series of windows as a sequence of time
evolutionary steps of a single tree or graph. Next we define a
distance between each pair of stocks, and base the distance on the
correlation coefficient. The transformation $d^t_{ij}=\sqrt{2(1-\rho
  _{ij}^t)}$ is motivated by considerations of ultrametricity
\cite{Man1}. For reasons of compatibility with the earlier work we
will use this definition, but would like to point out that for our
purposes any monotonically decreasing distance function of the
correlation coefficient $\rho _{ij}^t$ would do. With the chosen
transformation, the individual correlation coefficients are mapped
from $[-1,1]$ to $[2,0]$, and the correlation matrix is mapped into a
symmetric distance matrix $\mathbf{D}^t$. 

Until now the method for constructing asset trees and asset graphs is identical, 
and the difference arises in the next step. Asset trees are
constructed according to \cite{Man1} by determining the minimum
spanning tree (MST) of the distances, denoted $\mathbf{T}^t$. The
spanning tree is a simply connected acyclic graph that connects all
$N$ nodes (stocks) and its size (number of edges) is fixed at $N-1$
such that the sum of all edge weights, $\sum _{d_{ij}^t \in
\mathbf{T}^t}d_{ij}^t$, is minimum. The spanning tree, by definition,
spans all $N$ vertices in the set $V$ in all time windows $t$ and is
thus time independent, whereas the set of edges $E^t$ is time
dependent, as is evidenced by our studies on tree robustness in
\cite{short, long, bali}. In contrast, asset graphs are created for
the same set of vertices but the edges are inserted one by one,
according to the rank of the corresponding element of the $\mathbf
{D}$ matrix such that we start with the smallest (i.e., with the
highest correlation). Therefore the asset graph can have any size
between 0 and $N(N-1)/2$, corresponding to all vertices being isolated
and the entire graph being fully connected, respectively. The size $n$
is controlled by the number of shortest edges already
present in the graph.  There is no acyclicity condition for asset
graphs, neither do they need to be connected. 

\section{Asset graph and asset tree comparisons}

Let us now consider, as a special case, an asset graph of order $N$
(number of vertices or stocks), and of size $n=N-1$ (number if edges),
so that it is comparable in this sense to the asset tree. In general,
the elements included in the asset graph are much more optimal, i.e.,
shorter than those in the asset tree, as can be shown by examining
their distributions, see \cite{intia}. This is due to the fact that
there are very strongly inter-connected clusters in the market, and
they are reproduced in the asset graph, but not in the asset tree
where the tree condition suppresses this feature. Thus some of the
vertices form cliques, use up the available edges and create
cycles in the process. On the other hand, the spanning criterion
forces the tree to include weak connections which are naturally left out
from the graph. For a visualisation of these differences  see Figures 1 and 2 in \cite{intia}.
 
\begin{figure}
\resizebox{0.5\textwidth}{!}{%
  \includegraphics{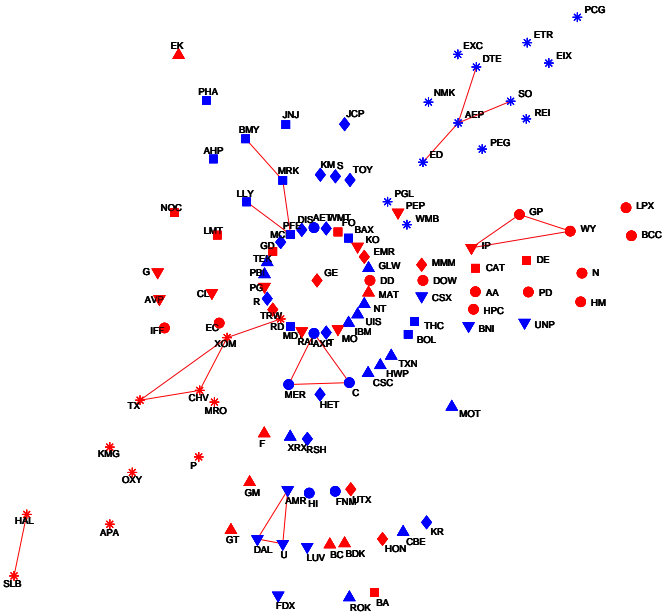}
}
\caption{Sample graph of $N=116$ vertices and $n=20$ edges, corresponding to a connection probability $p=n/[N(N-1)/2] \approx 0.003$.}
\label{graafi1}       
\end{figure}

\begin{figure}
\resizebox{0.5\textwidth}{!}{%
  \includegraphics{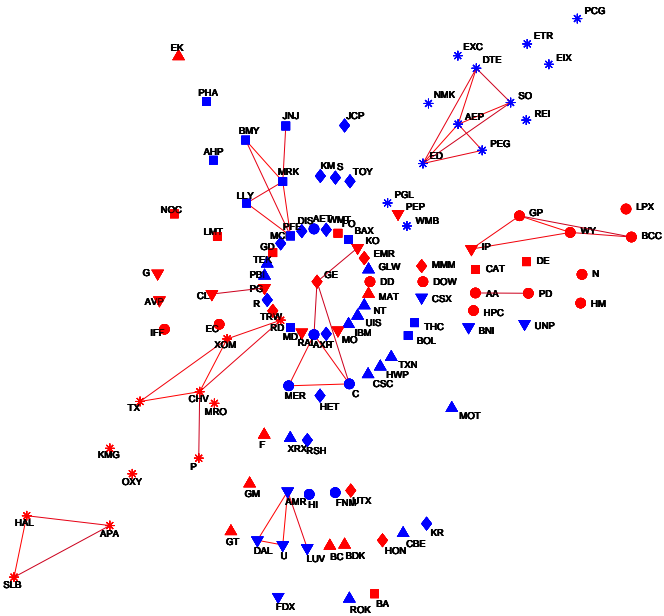}
}
\caption{Sample graph for $n=40$ edges ($p \approx 0.006$).}
\label{graafi2}       
\end{figure}

\begin{figure}
\resizebox{0.5\textwidth}{!}{%
  \includegraphics{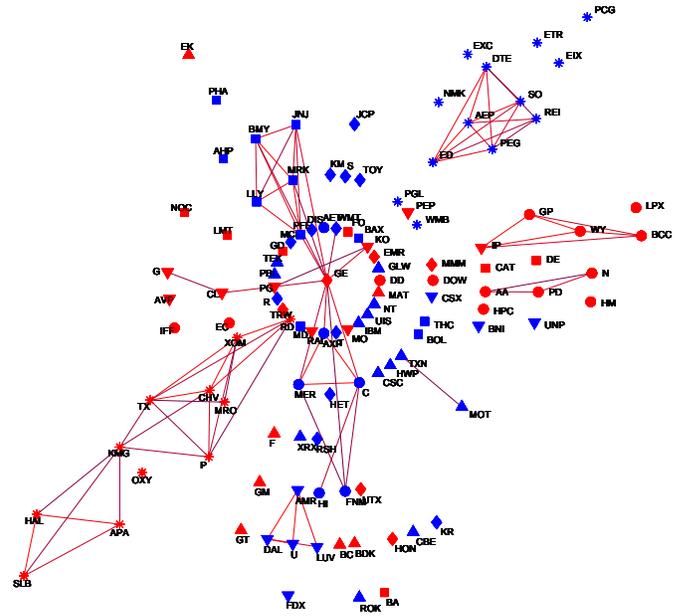}
}
\caption{Sample graph for $n=80$ edges ($p \approx 0.012$).}
\label{graafi3}       
\end{figure}

\begin{figure}
\resizebox{0.5\textwidth}{!}{%
  \includegraphics{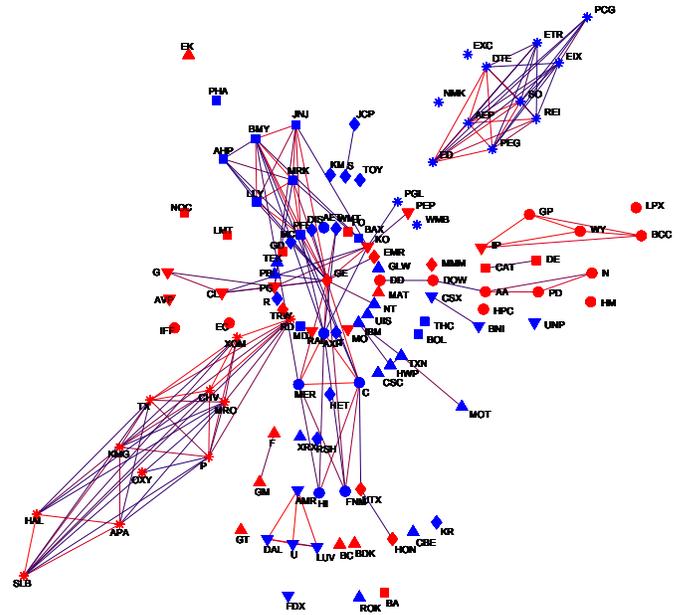}
}
\caption{Sample graph for $n=160$ edges ($p \approx 0.024$).}
\label{graafi4}       
\end{figure}

Here we wish to focus more on the aspects of the growth and clustering for
the same set of data, in particular for the asset graph. The most
straight-forward way to see how the asset graph topology and clusters
form is depicted as an example in Figures \ref{graafi1} to
\ref{graafi4}. Note that vertices are drawn using a variety of
different markers, where the marker type and colour 
correspond to the company's business sector as classified by Forbes
\cite{forbes}. For certain companies, such as those in the Energy
Sector (marked by red asterisks) 
we would expect strong \emph{intra}-business sector clustering, and
for some, such as those in the Financial business sector (blue
circles), 
we would expect strong \emph{inter}-business sector clustering. There
are also some stocks for which we would not expect graph clustering to
correspond to the business sector labels (for a discussion on the
correspondence between business sectors and asset tree clusters see
\cite{long}).

Some observations and comments are in place. 

(i) In Figure
\ref{graafi1}, after only $n=20$ edges have been added, already four
cycles have formed. This makes it clear that asset tree and asset
graph topologies start to diverge at an early stage, i.e., for small
$n$. 

(ii) In Figure \ref{graafi2}, the additional 20 edges seem to
reinforce the small clusters present in Figure \ref{graafi1}. In
general, it is interesting to note that the clusters created very
early seem to become more and more strongly connected, and also grow
by having new vertices attached to them as edges are added. It is not
evident that the strongest connections (shortest edges) should define
the clusters the way they do, as one could have a situation where
a very strongly cliqued group of companies appears later on. However,
moving from Figure \ref{graafi2} to \ref{graafi4}, it is clear that this is what happens. 

(iii) An asset tree defined on
116 vertices has 115 edges. In Figure \ref{graafi4}, where the
number of edges $n=160$ easily exceeds this, there are still several
isolated vertices left. This turns out to be so even after 1000 edges
have been added. The asset tree, however, would contain by definition
those isolated vertices after the inclusion of $n=115$ edges. In this
sense, although the asset tree can provide an overall taxonomy of the
market, the connections it creates may be misinterpreted to be more
meaningful than they are. As mentioned earlier and studied in
\cite{intia}, this due to the the minimum spanning tree
criterion. Consequently, it is hardly surprising that an asset graph
of the size of an asset tree is much more robust, since the weak
connections contained in the tree are prone to breaking easily
\cite{intia}.  

(iv) We can observe in Figure \ref{graafi4} that although
some clusters are very heavily intra-connected, they are not yet inter-connected 
to other clusters. Two such examples are the energy cluster
at the bottom left corner and the utilities cluster in the top right
corner of Figure \ref{graafi4}. 

(v) In general, we see that there is
good agreement between graph clusters and business sector definitions
given by an outside institution. 

(vi) Although the graph analysed here
is just a sample, obtained by fixing the time, i.e., choosing a random
value for the time superscript $t$, preliminary studies indicate that
qualitatively similar clustering is observed throughout the time
domain.

\begin{figure}
\resizebox{0.5\textwidth}{!}{%
  \includegraphics{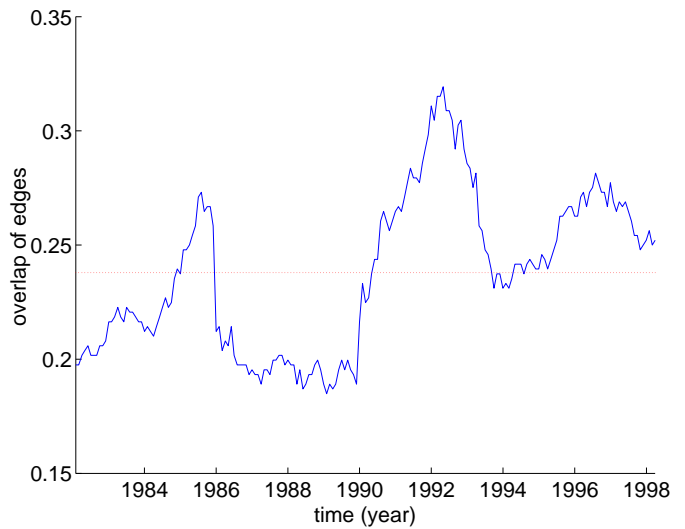}
}
\caption{Overlap of edges in the asset graph $\mathbf{G}^t$ and 
  asset tree $\mathbf{T}^t$ for $T=1000$ trading days as a function of time. 
The average value,  roughly 24\%, is indicated by the horizontal line.} 
\label{overlap_vs_time}
\end{figure}

As points (i) and (iii) above indicate, asset trees and asset graphs
have clearly different topologies. Let us denote the asset graph more
completely by its vertex and edge set as
$\mathbf{G}^t=(V_G,E_G^t)$, and the asset tree similarly by
$\mathbf{T}^t=(V_T,E_T^t)$. For statistically more reliable results, 
we have used a set of split-adjusted daily price data for $N=477$ NYSE
traded stocks, time-wise extending from the beginning of 1980 to the
end of 1999. This is the dataset we will use throughout the paper
unless mentioned otherwise. We can learn about the overall topological
differences between the asset graph and asset tree by studying the
overlap of edges present in both as a function of time. The relative
overlap is given by $\frac{1}{N-1}|E_G^t \cap   E_T^t|$ where $\cap $
is the intersection operator and $|...|$ gives the number of elements
in the set. As can be see from the plot in Figure
\ref{overlap_vs_time}, on average the asset graph and asset tree share
about 24\%, or roughly one quarter, of edges. This quantity is also
fairly stable over time. Since the asset graph consists of the
shortest possible edges and is optimal in this sense, whenever an edge
in $E_T^t$ is not included in $E_G^t$, the sum of edges for the asset
graph is increased above this optimum. Therefore, we can infer from Figure
\ref{overlap_vs_time} that on average some 75\% of the edges contained
in the asset tree are not optimal in this sense. 
We drew a similar conclusion by comparing edge length distributions   
for the asset tree and asset graph in Figures 4 and 5 \cite{intia}.

\begin{figure}
\resizebox{0.5\textwidth}{!}{%
  \includegraphics{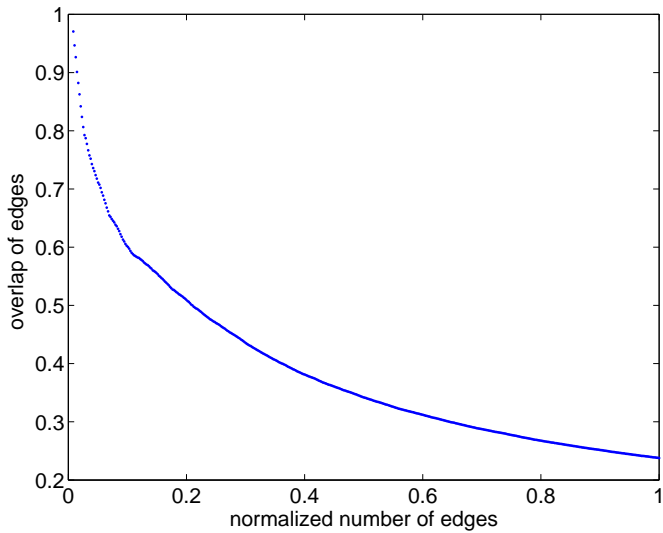}
}
\caption{Overlap of edges $E_G^t(n)$ in the asset graph and $E_T^t(n)$ in the asset tree, where $n=1,2,\ldots, n$, as a function of normalised number of edges $\frac{n}{N-1}$, averaged over time.}
\label{overlap_vs_edges}
\end{figure}

Motivated by observation (i) above, it is also of interest to study
how this overlap of edges changes in the process of constructing asset
graph and tree one edge at a time. In order to generate the minimum
spanning tree, we use Kruskal's algorithm. This consists of taking all
of the distinct $N(N-1)/2$ distance elements from the distance matrix
$\mathbf{D}^t$, and obtaining a sequence of edges $d_{1}^t, d_{2}^t,
\ldots, d_{N(N-1)/2}^t$, where we have used a single index
notation. The edges are then sorted in a nondecreasing order to get an
ordered sequence $d_{(1)}^t, d_{(2)}^t, \ldots, d_{(N(N-1)/2)}^t$. We
select the shortest unexamined edge for inclusion in the tree, with
the condition that it does not form a cycle. If it does, we discard
it, and move on to the next unexamined edge on the list. Apart from
for the constraint on cycles, the algorithm is identical to the way
asset graphs are generated. If we denote the
size of graph in construction by $n$, where $n=1,2, \ldots, N-1$, then
at least for small values of $n$ asset graphs and asset trees should
contain the same set of edges, i.e., $E_G^t(n) = E_T^t(n)$ and,
therefore, be identical in topology. It is expected that, starting
from some value of $n=n_c$, the above equality no longer holds, and 
observation (i) above leads us to expect a small value for $n_c$. 
Once the equality breaks, the first cycle is formed and,
consequently, for all $n \ge n_c$ the asset graph and tree differ
topologically. This is demonstrated in Figure \ref{overlap_vs_edges},
where the relative overlap of edges, $\frac{1}{N-1}|E_G^t(n) \cap E_T^t(n)|$, 
has been plotted as a function of normalised number of edges,
$\frac{n}{N-1}$, and the quantity has been averaged over time. The
function decreases rapidly for small values of $\frac{n}{N-1}$,
indicating that for the current set of data with $N=477$, only a few
edges can be added before the first cycle is formed. As more and more
edges are added, the plot converges to the 24\% time average.

\section{Asset graph and random graph comparisons}

We now leave asset trees behind and deal exclusively with asset graphs. 
We focus on our empirical sample
graph $\mathbf{G}_{\text{emp}}$ evaluated from a distance matrix
$\mathbf{D}^t$ for a randomly chosen time window location $t$. We then
construct a random graph of the same size as the asset graph, and
compare the results between the two. The fact the window is fairly
wide at $T=1000$ means that the results are less sensitive to the
time location $t$ of the window and, consequently, can be generalised to a
greater extent than if a shorter window width was used. Time
dependence of the quantities studied, as well as a more analytical
approach in general, are postponed until a later exposition.  

As should be clear from the earlier discussion, the asset tree
approach as a simple, non-parametric classification scheme always
produces a unique taxonomy. Because of the tree condition, the asset
tree ignores some important correlations, and also fails to capture
the strong networking present in the financial market. It is generally
agreed that the correlation matrix contains both information and
noise, and one is obviously interested in finding and studying the
information rich part. In the extreme case of no information, one
could find the minimum spanning tree for a completely random matrix of
uncorrelated data. In this case one would also obtain a
classification, but hardly a meaningful one. This indicates a possible drawback
in the minimum spanning tree methodology.  

Growth and clustering of asset graphs is an interesting problem in its
own right, but it may also, as we believe, shed light on the
information versus noise issue. We will now consider the size $n$ of
the graph as a parameter and increase it, at least in theory, all
the way up to the fully connected graph. If $d_{(n)}$ is the latest
edge added, where $n=1,2, \ldots, N(N-1)/2$, we quantify the degree of
graph completeness by $p=n/[N(N-1)/2]$, where $p \in [0,1]$. In
practice, for our empirical data of $N=477$ stocks we do this for $p
\in [0, 0.25]$, corresponding to a maximum of \mbox{28,382} edges. In
our experience this interval is sufficient, since most quantities
beyond this become practically random anyway. 

The random graph, or more specifically an Erdös-Rényi random graph, is
denoted by $\mathbf{G}_{\text{ran}}$ and constructed as follows: Given
$N$ labelled, isolated vertices, we consider all possible vertex pairs
in turn and connect them with probability $p$. However,
instead of generating the random graph explicitly from the definition,
we obtain one by shuffling the elements in the distance matrix
$\mathbf{D}^t$ and then add them, one edge at a time, to the
graph. The graphs obtained at different stages of this process correspond to higher and
higher connection probabilities $p$. This method enables us to compare
graph construction for the empirical graph $\mathbf{G}(p)_\text{emp}$
and random graph $\mathbf{G}(p)_\text{ran}$ as a function of the
connection probability $p$. Strictly speaking the results derived from
the random-graph theory apply only in the limit when the number of
nodes $N$ tends to infinity. Although the datasets we have studied
have either $N=116$ or $N=477$, acknowledging the presence of finite
size effects, one can consider the random graph as a benchmark against
which deviations from random behaviour can be measured. As we will
see, the financial network does not follow the predictions of the
random graph theory and thus constitutes a complex network.

\begin{figure}
\resizebox{0.5\textwidth}{!}{%
  \includegraphics{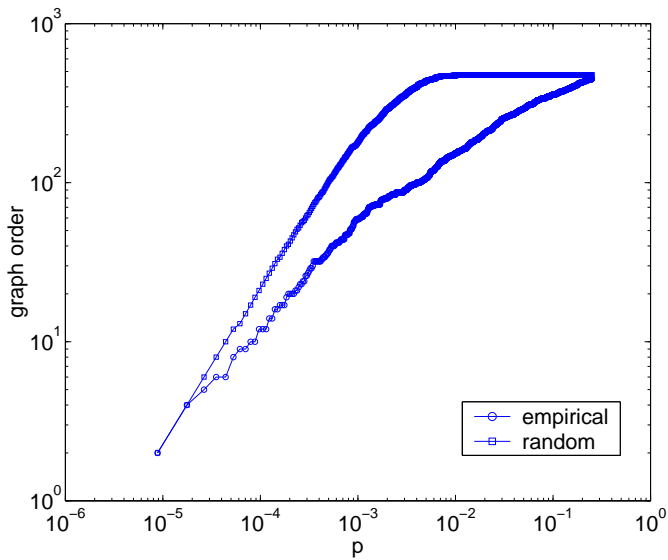}
}
\caption{Spanned graph order for empirical and random data.}
\label{graph_orders}       
\end{figure}

\subsection{Cluster growth and size}

We start by studying what we call the \emph{spanned graph
  order}. Whereas graph order indicates the number of vertices in the
graph, we define spanned graph order as the number of vertices with
vertex degrees greater than or equal to one, i.e., only those vertices
are counted that have at least one edge connected to them. This
distinction is needed because graph order itself is a constant for our
graphs. Figure \ref{graph_orders} plots spanned graph order for
empirical and random data. We find that the random graph becomes fully
connected very early on, i.e., its spanned graph order
$\mathcal{S}(\mathbf{G_\text{ran}}(p'))=N=477$ for $p' \approx 0.012$, whereas
for the empirical graph for the same value of $p$ we have
$\mathcal{S}(\mathbf{G}_\text{emp}(p'))=164$. In the empirical case, edges are
used to create strong clusters and, therefore, the spanned graph order
grows more slowly than for the random case, in which there is no
systematic clustering present. 


We can study some topological aspects of graph construction by
considering four distinct types of growth that occur in the
graphs. The division into these specific growth types is motivated by their
intuitive appeal and relevance in this application context. These
different types cause qualitatively different growth of graph
clusters, and studying them can help us understand 
the differences we observe in greater detail. In the case of a financial network, edge
clusters are more interesting than vertex clusters, because it is
edges, i.e., correlations amongst stocks, that very naturally define
clusters in the financial market, as Figures \ref{graafi1} to
\ref{graafi4} show. A \emph{cluster}, denoted by
$\mathcal{C}_i=(V_i,E_i)$, is defined to be an isolated subgraph
induced by a set of edges $E_i$, containing the vertices $V_i$. We
also define \emph{cluster size} of $\mathcal{C}_i$ simply as
$|E_i|$. Similarly, \emph{cluster order} for $\mathcal{C}_i$ is given
by $|V_i|$. The four different growth types occurring upon the addition of a new edge $e_{ij}$, 
incident on vertices $v_i$ and $v_j$, are the following:

\newcounter{bean}
\begin{list}
{(\Roman{bean})} {\usecounter{bean} \setlength{\rightmargin}{\leftmargin}}
\item \emph{Create a new cluster.} This occurs when neither of the two
  vertices $v_i$ nor $v_j$, incident on the new edge $e_{ij}$, are part of
  an existing cluster. A new cluster is created, its spanned cluster
  order is two, and cluster size one. 
\item \emph{Add a node and an edge to an existing cluster.} Adds
  vertex $v_i$ and the incident edge $e_{ij}$ to an existing cluster,
  when the other vertex $v_j$ already belongs to it. Spanned cluster
  order and cluster size are increased by one. 
\item \emph{Merge two clusters.} Merge cluster $\mathcal{C}_i$
  containing the vertex $v_i$ and cluster $\mathcal{C}_j$ containing
  the vertex $v_j$ by adding the incident edge $e{_{ij}}$ between
  them. If $|E_i| \ge |E_j|$, the cluster $\mathcal{C}_i$ survives and
  its new order is $|V_i| + |V_j|$ and new size $|E_i| + |E_j| +
  1$. Cluster $\mathcal{C}_j$ disappears as we have $E_j =
  \emptyset$ and $V_j = \emptyset$. Intuitively speaking, the larger
  cluster eats the smaller one.  
\item \emph{Add a cycle to an existing cluster.} Add an edge to an
  existing cluster, thus creating a cycle and reinforcing the
  clustering. Spanned graph order is increased by one. 
\end{list}

\begin{figure}
\resizebox{0.5\textwidth}{!}{%
  \includegraphics{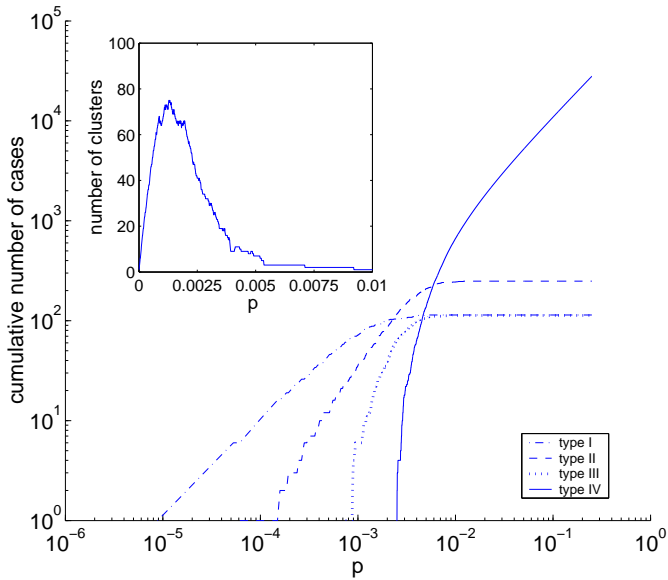}
}
\caption{Growth types for the random graph. Inset: number of clusters for the random graph.}
\label{growth_ran}       
\end{figure}

\begin{figure}
\resizebox{0.5\textwidth}{!}{%
  \includegraphics{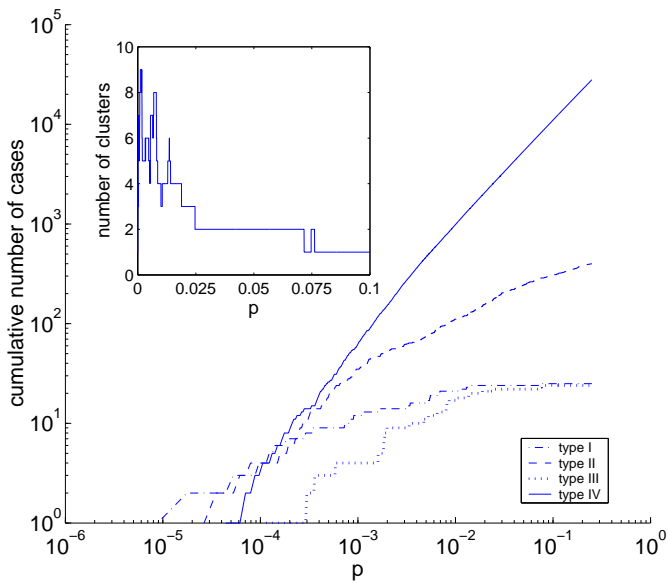}
}
\caption{Growth types for the empirical graph. Inset: Number of clusters for the empirical graph.}
\label{growth_emp}       
\end{figure}

The cumulative occurrence of each growth type is plotted as a function of $p$ for 
random data in Figure \ref{growth_ran} and for empirical data in
Figure \ref{growth_emp}. Some observations. (i) The growth of the
random graph starts linearly with type I and continues like that
practically for two decades, as new clusters of one edge and two
vertices are created. As a result, the number of vertices grows by two
on each step, contributing to the rapid increase in spanned graph
order for the random graph in Figure \ref{graph_orders}. Type I growth
is clearly less dominant for the empirical graph, for which growth of
other types starts earlier. (ii) In regard to clustering, type IV
growth is most relevant and is observed roughly 1.5 decades earlier
for the empirical data than for the random data. This finding is
corroborated by Figures \ref{graafi1} to \ref{graafi4} and the related
discussion. (iii) We observe that the number of types I and III growth
almost converge as $p \to 1$. The convergence is to be expected since
in moving towards a fully connected graph, all the separate clusters
that have been formed will be merged at some point. Thus in the limit the number
of mergers needs to equal the number of components to be merged minus
one, since one cluster, the fully connected graph, remains. The
convergence seems to take place an estimated 1.5 decades later for the
empirical graph than for the random graph, indicating that the clusters
observed for the empirical data remain separate or disconnected from
the rest until much later.  


Let us now study the number of clusters formed as a function of
$p$. Of the four growth types analysed above, only type I and type III
affect the number of clusters in the system, by either increasing or
decreasing it by one, respectively. Therefore, the number of clusters
for a given value of $p$ is given by the difference between type I and type III
curves in Figures \ref{growth_ran} and \ref{growth_emp}. This is more
clearly shown on linear scales in the insets of the same figures 
(please note that the scales in the insets are different). 
The maximum number of clusters for the sample random graph is 75,
occurring at $p \approx 0.0013$, whereas for the empirical graph it is
9, occurring at $p \approx 0.0011$. The high spanned graph order for
the random graph due to type I growth, and relatively low mean
clustering coefficient as compared to the asset graph (as seen later),
leads to a large number of clusters that are relatively early combined
to form one giant cluster. In contrast, the empirical graph has a much
more slowly increasing spanned graph order, fewer clusters, and
exhibits predominantly type IV growth to enhance the existing clusters
(high mean clustering). Consequently, the maximum number of clusters
is left small. It is interesting to note that in this case the maxima,
although very different in value, happen for roughly the same value of
$p$. Further studies are required to explain whether this is by chance
or a systematic finding. 

\begin{figure}
\resizebox{0.5\textwidth}{!}{%
  \includegraphics{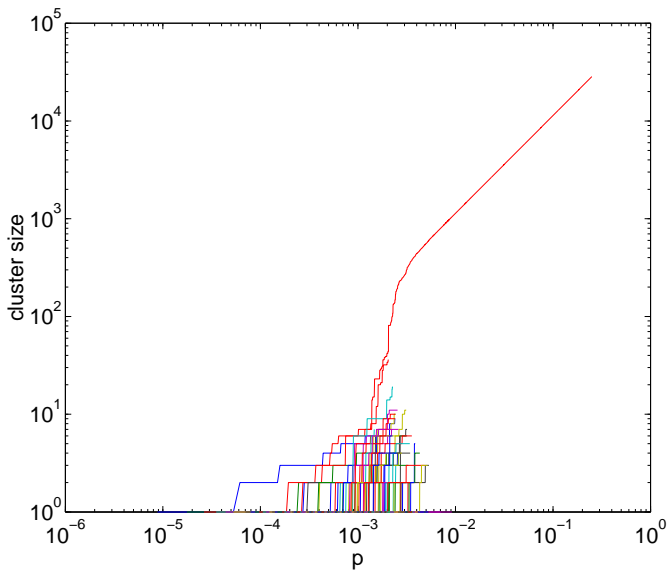}
}
\caption{Cluster size for the random graph. Different curves correspond to
  different clusters. Since several clusters of size one overlap one
  another in this figure rendering them indistinguishable, one cannot
  count the total number of clusters from this plot.} 
\label{cs_ran}       
\end{figure}

\begin{figure}
\resizebox{0.5\textwidth}{!}{%
  \includegraphics{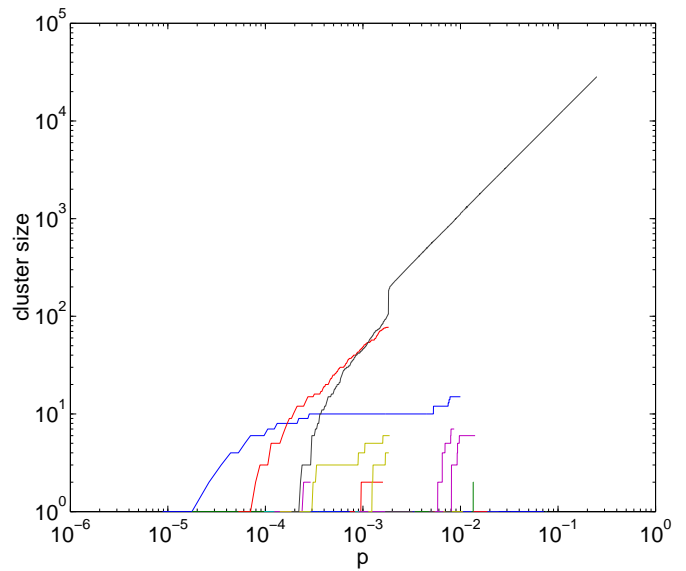}
}
\caption{Cluster size for the empirical graph. See comment in Figure \ref{cs_ran}.}
\label{cs_emp}       
\end{figure}

Let us now turn to cluster size distributions presented in Figures
\ref{cs_ran} and \ref{cs_emp}. For the random graph, the large number
of clusters seem to disappear suddenly when the clusters are merged
together, as the sudden jump in type III growth in Figure
\ref{growth_ran} indicates. This type of sudden transition is not
present for the empirical graph, further supporting the conjecture
that the behaviour of the asset graph is markedly different from the
random graph.  

The results we have obtained for the random graph are well explained
by some basic random graph theory, from which we wish to review very
briefly some important elementary findings \cite{barabasi}. 
This will help not only to explain the random results, but may also help to understand why the 
empirical graph behaves so differently. The most central goal of
random-graph theory is to determine at what connection probability $p$
a particular property of a graph will most likely arise. In most
general terms, we can ask whether there is a critical probability that
marks the appearance of arbitrary subgraphs and, as its important
special cases, trees and cycles of a given order. The problem was
solved by Bollobás \cite{bollobas}. Consider a random graph with $N$
vertices connected by $n$ edges and assume that the connection
probability $p(N) \propto N^z$, where the parameter $z \in (- \infty,
0]$. For a random graph, the average degree is given by 
\[
\langle k \rangle  = 2n/N = p(N-1) \approx pN,
\]

\noindent and this quantity has a system size independent critical
value. When $z < -1$ such that the average degree of the graph $
\langle k \rangle = pN \to 0$ as $N \to \infty$, the graph consists of
disjoint trees. The appearance of these small trees is tied to some
threshold values of $z$ such that below that value almost no graph has
the given property, whereas for values above it almost every graph has
the property. What is remarkable from our perspective is that for $z <
-1$ there are no cycles present, but when $z=-1$, corresponding to
$\langle k \rangle = $ constant, trees and cycles of all orders
appear. We can find out about the size and structure of clusters for
this particular case when $p \propto N^{-1}$. When $0 < \langle k
\rangle < 1$, although there are cycles present, almost all nodes
belong to trees, and the size of the largest tree is proportional to
$\ln N$. The mean number of clusters is of order $N-n$, so in this
range of $\langle k \rangle$ the number of clusters decreases by 1 as
$n$ increases by 1, i.e., when a new edge is introduced in the
graph. If $\langle k \rangle$ is increased to the threshold $\langle k
\rangle _c = 1$, corresponding to a critical probability $p_c \approx
1/N$, the topology of the graph changes suddenly. The small clusters
are merged together to form a single giant cluster, or a giant
component, and it has a fairly complex structure. Other clusters are
small, and most of them are trees. As $\langle k \rangle$ is increased
further, the small clusters are attached to the giant
cluster. Therefore, for values below $p_c$ the graph is made up of
isolated clusters, but for values above $p_c$ the giant cluster spans
the graph. Given these theoretical considerations, the fact that
cycles are found in the graphs in Figures \ref{graafi1} to
\ref{graafi4}, even for $p \approx 0.003$, underlines the highly
correlated ``non-random'' nature of the financial network. 
Last, as a point concerning terminology, it should be
mentioned that the emergence of the giant cluster is the same
phenomenon as a percolation transition in infinite-dimensional
(mean field) percolation. The difference in the
behaviour around the emergence of the giant component between the
random and empirical graph indicates that the transition in the
latter is also of different nature.

\subsection{Clustering coefficients and information}

\begin{figure}
\resizebox{0.5\textwidth}{!}{%
  \includegraphics{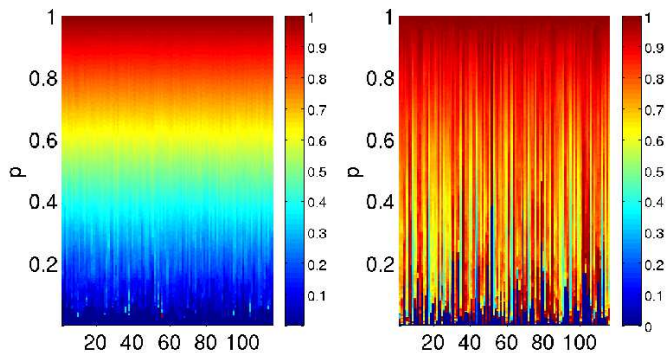}
}
\caption{Clustering coefficient as a function of vertex index (horizontal axis) and $p$ (vertical axis). Left: random graph, right: empirical graph.}
\label{vertex_ccs}       
\end{figure}

Finally, we will study the clustering coefficients for our smaller set
of 116 S\&P500 stocks. Clustering coefficient of vertex $i$ is defined
as 

\[
C_i = \frac{2 \Delta_i}{k_i(k_i-1)},
\]

\noindent where $k_i$ is the number of incident edges of vertex $v_i$ (vertex degree), 
and $\Delta_i$ the number of edges that exist between the
$k_i$ neighbours of vertex $v_i$. The normalisation in the definition is
due to the fact that at most there can be $k_i(k_i-1)/2$ edges between
the $k_i$ vertices, which would happen if they formed a fully
connected subgraph. Thus the coefficient is normalised on the interval
$[0,1]$. The value of clustering coefficient for each vertex $v_1,
v_2, \ldots , v_{116}$ is plotted in Figure \ref{vertex_ccs} for both
the random graph and empirical graph, where the vertex index is given
on the horizontal axes, the vertical axes give the value of $p$, and
the shades 
corresponds to the value of the clustering coefficient. The two plots
are strikingly different. For the random graph, overall there is a
very smooth, rainbow-like transition from zero to unity. In addition,
all vertices behave in a fairly homogeneous manner. For the empirical
graph the transition towards unity is much faster and there is much
greater heterogeneity present. Further, there are some very high clustering
coefficient values observed for some vertices at low values of $p$.  

Since much of our attention has focused on asset graph clusters, we calculated clustering coefficients of the sample graph for each cluster when $p \in [0,1]$. These are simply averages of the clustering coefficients $C_i$ of individual vertices belonging to a given cluster $\mathcal{C}_i$, i.e., 
\[
C_{\mathcal{C}_i}= \frac{1}{|V_i|} \sum_{C_i \in \mathcal{C}_i }C_i.
\]

\begin{figure}
\resizebox{0.5\textwidth}{!}{%
  \includegraphics{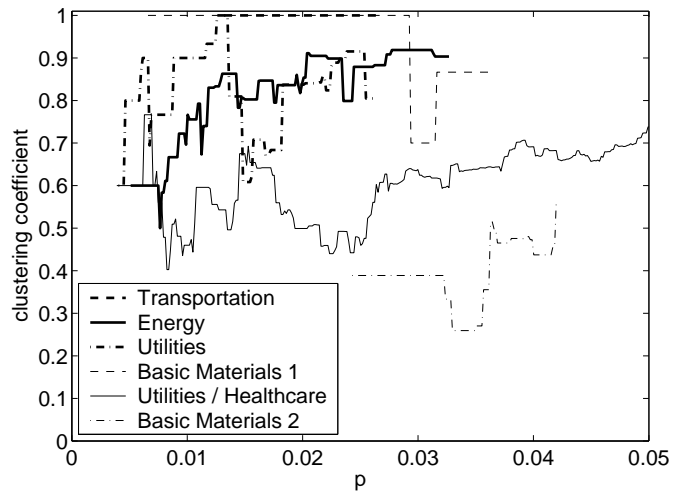}
}
\caption{Clustering coefficients for selected clusters as a function of $p$.}
\label{cluster_ccs}       
\end{figure}

\noindent In Figure \ref{cluster_ccs} we show results for selected six clusters, namely, Transportation, Energy, Utilities, Basic Materials 1, Utilities / Healthcare, and Basic Materials 2. For values of $p \ge 0.05$ all other clusters coalesce into the Utilities / Healthcare cluster, which behaves very similarly to the mean clustering coefficient discussed shortly. The small deviations result from the fact that there are some isolated vertices which are not included in the coalesced cluster but are counted in the mean clustering coefficient. For purposes of visualisation only clusters with six or more edges are included in Figure \ref{cluster_ccs}, as for smaller clusters the clustering coefficient fluctuates wildly and makes the plot messy. Further, only those clusters with reasonably long life time in terms of $p$ are included. 

In most cases each cluster consists of stocks that belong to different business sectors. The clusters are named after the dominating business sector, i.e., the business sector shared by a majority of the vertices in the cluster. Apart from one exception, a single business sector dominates for each value of $p$, indicating strong correspondence between cluster and business sector groups. The only exception is the largest cluster, i.e., Utilities / Healthcare, which was dominated by either Utilities or Healthcare stocks, depending on the value of $p$.

The four most highly connected clusters are Transportation, Basic Materials 1, Utilities, and Energy. The cluster-wise calculated clustering coefficients are more meaningful when examined in conjunction with Figures \ref{graafi1} to \ref{graafi4}. One should also bear in mind that cluster sizes and cluster orders for the four clusters are different, and this needs to be taken into account when studying clustering coefficients. Although cluster sizes for these clusters are not reported in this paper for the particular set of data, it is clear that for larger clusters there is more jitter in the curves of Figure \ref{cluster_ccs}. The Transportation cluster consists, for the most part, of stocks AMR, DAL, U and LUV and is fully connected, as there is an edge between DAL and LUV, although poorly visible. Basic Materials 1 cluster consists of stocks IP, GP, WY and BCC, and they are also fully connected for $p \in [0.005, 0.03]$, but clustering falls as new vertex is added to the cluster. The most striking examples, however, are Utilities and Energy clusters, both of which encompass several vertices. As Figure \ref{graafi4} shows, they are very strongly connected. Quite remarkably, both clusters are also very homogeneous in terms of their business sector makeup. These findings indicate that in the financial network there are clusters that are relatively separate from others, and yet their internal connectivity is high.

\begin{figure}
\resizebox{0.5\textwidth}{!}{%
  \includegraphics{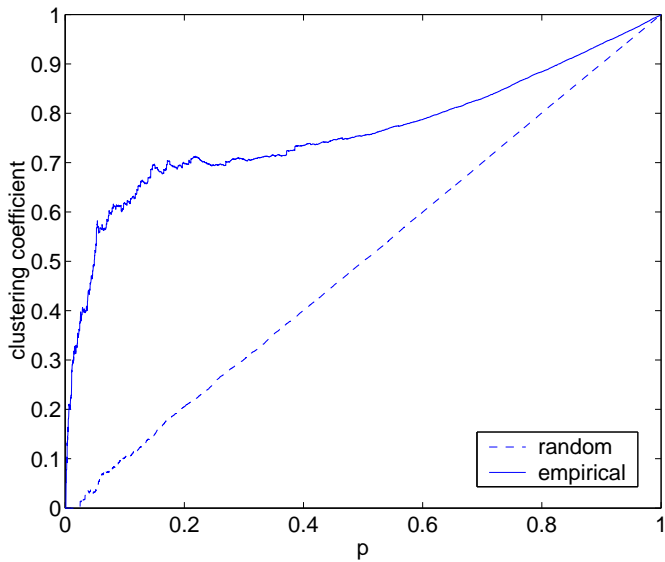}
}
\caption{Mean clustering coefficients for the random and empirical graph as a function of $p$.}
\label{mean_ccs}       
\end{figure}

By averaging the clustering coefficients $C_i$ over all vertices $i$
one obtains the \emph{mean clustering coefficient} $C_{\text{ran}}$
and $C_{\text{emp}}$, both plotted in Figure \ref{mean_ccs}. From this
plot the difference in the rate of change of the clustering
coefficient for the random and empirical case is very obvious. For the
random graph the mean clustering coefficient is zero up to and
including $p' = 125/6670 \approx 0.02$, whereas for the empirical
graph for the same $p=p'$ the mean clustering coefficient is 0.33. For
the random graph, the zero value and low values at the beginning in
general are again explained by type I growth leading to duple
clusters (one edge, two vertices), for which the clustering
coefficient is zero. For the empirical graph the early type IV growth
creates several cycles of order three as can be seen, for example, in
Figure \ref{graafi1}. For these cycles the clustering coefficient is
unity, and this contributes to the mean clustering coefficient. To
visualise the empirical graph with 125 edges, one can mentally
interpolate between Figures \ref{graafi3} and \ref{graafi4} to
convince himself or herself of the high mean clustering coefficient
value. Please note that the clustering coefficient results can
directly be compared only with Figures \ref{graafi1} to \ref{graafi4},
since for other random and empirical graph plots a different dataset
was used.  


The mean clustering coefficient for the random graph, for all
practical purposes, is linear with a slope of unity (except for the
slight fluctuation for small $p$). This result is compatible with
random graph theory, since for a random network, the probability of
its two nearest neighbours being connected is the same as that for any
two randomly picked vertices being connected. Therefore, the mean
clustering coefficient for a random graph is  

\[
C_{\text{ran}}=p=\frac{\langle k \rangle}{N}.
\]

We conjecture that comparing the mean clustering coefficient of an
empirical asset graph against a random graph can be used to estimate
the information content of the edges in the graph and, consequently,
the information content of the corresponding correlation coefficients
in the related correlation matrix. For a rough analysis of results we
divide the empirical curve in Figure \ref{mean_ccs}, based on its
behaviour, into three sections along the horizontal axis. The first
section of rapid growth covers the first 10\% of edges ($p \in
[0,0.1]$), during which the mean clustering coefficient increases very
rapidly and, in particular, much faster than for the random graph. We
interpret this significant deviation from the random case to imply
that the first 10\% of the edges add substantial information to the 
system. During the first part of the second section for roughly $p \in [0.1,0.2]$, 
the rate of change starts to slow down and reaches a sort
of a plateau or saturation during the second part of this section for
$p \in [0.2,0.3]$. We consider these findings to indicate that the edges added
in this section for $p \in [0.1,0.3]$ are less informative. For the last section, from
$p=0.3$ onwards, we believe the remaining 70\% to be relatively poor
in information content, possibly just noise. Although the curve
becomes steeper as $p \to 1$, we do not consider this to reflect
genuine information but to result from the boundary conditions of the
problem, since for $p=1$ the mean clustering coefficient must be equal to unity.  

We believe that the method of comparing empirical graph properties to
random graph theory predictions can be used to address the information
versus noise issue of the underlying correlation matrix. In spirit
this is a similar argument to using random matrix theory to study the
information content of empirical correlation matrices by comparing
their properties, mainly eigenvalue spectra. In \cite{Lal}, there was
remarkable agreement between the theoretical prediction and empirical
data concerning both the density of eigenvalues and the structure of
eigenvectors for the correlation matrix. For their set of $N=406$
assets of the S\&P 500 for $T=1309$ days, Laloux \emph{et al} found 94\% of
the total number of eigenvalues to fall within the region predicted by
the theory, leaving only 6\% of the eigenvectors to appear to carry
some information. This finding is compatible with the above discussion. 
We plan to repeat this analysis for a larger set of data in the
near future and carry it out dynamically.

\vskip .3in
\vskip .3in

\section{Summary and Conclusion}

In this paper we have recapitulated the methodology for constructing
asset graphs and asset trees. Due to the tree condition, the asset
tree fails to capture the strong clustering in the financial market,
but this is clearly present in the asset graph. We have found the
clusters in the asset graph to appear very early, i.e., for low
connection probabilities, after which asset graph and asset tree topologies begin to differ. The two
methodologies result in an approximate 25\% overlap of edges over
time, and the remaining 75\% cause them to exhibit qualitatively very
different behaviour. We have studied the asset graph further and
compared the results to a random graph of the same size as a function
of connection probability. We have divided the growth processes into
four distinct growth types, and have found type I growth to be
responsible for the fast growth in spanned graph order for the random
graph. A study of growth types has also revealed how type IV growth,
responsible for creating cycles in the graph, sets in much earlier for
the asset graph, and thus reflects the networking present in the
market. We have also found the number of clusters in the random graph
to be one order of magnitude higher than for the asset graph. At a
critical threshold, the random graph undergoes a radical change in
topology, when the small clusters merge to form a single giant
cluster. This phenomenon, equivalent to a percolation transition, is
not observed for the asset graph. Finally, we have studied clustering
coefficients and mean clustering coefficients, and found them to
behave very differently for the asset and random graph. We have
conjectured that this difference may be suitable for studying what
fraction of edges in the graph, or correlation coefficients in the
related correlation matrix, is information and what is noise. Based on
this approach, only some 10\% of the edges appear to carry genuine information. 
The results presented in this paper concerning asset and
random graph comparisons have been carried out for a randomly selected
but representative time window and a more rigorous study should be
made to include the possible effects of time dependence.

\noindent \textbf{Acknowledgements}

\noindent 
We are thankful to A. Chakraborti who participated in earlier stages of this work.
J.-P. O. is grateful to the Graduate School in Computational Methods
of Information Technology (ComMIT), Finland. This research was
partially supported by the Academy of Finland, Research Centre for
Computational Science and Engineering, project no. 44897 (Finnish
Centre of Excellence Programme 2000-2005).

\end{document}